\documentstyle[12pt]{article}

\advance \topmargin by -\headheight
\advance \topmargin by -\headsep
     
\evensidemargin \oddsidemargin
\begin{document}
\newcommand{\sect}[1]{\setcounter{equation}{0}\section{#1}}
\renewcommand{\theequation}{\thesection.\arabic{equation}}

\topmargin -.6in
\def\nonu{\nonumber}
\def\rf#1{(\ref{eq:#1})}
\def\lab#1{\label{eq:#1}} 
\def\br{\begin{eqnarray}}
\def\er{\end{eqnarray}}
\def\be{\begin{equation}}
\def\ee{\end{equation}}
\def\0{\nonumber}
\def\lb{\lbrack}
\def\rb{\rbrack}
\def\({\left(}
\def\){\right)}
\def\v{\vert}
\def\bv{\bigm\vert}
\def\lskip{\vskip\baselineskip\vskip-\parskip\noindent}
\relax
\newcommand{\nit}{\noindent}
\newcommand{\ct}[1]{\cite{#1}}
\newcommand{\bi}[1]{\bibitem{#1}}
\def\a{\alpha}
\def\b{\beta}
\def\ca{{\cal A}}
\def\cm{{\cal M}}
\def\cn{{\cal N}}
\def\cf{{\cal F}}
\def\d{\delta}
\def\D{\Delta}
\def\eps{\epsilon}
\def\g{\gamma}
\def\G{\Gamma}
\def\grad{\nabla}
\def\h{ {1\over 2}  }
\def\hc{\hat{c}}
\def\hd{\hat{d}}
\def\hg{\hat{g}}
\def\hp{ {+{1\over 2}}  }
\def\hm{ {-{1\over 2}}  }
\def\k{\kappa}
\def\l{\lambda}
\def\L{\Lambda}
\def\lg{\langle}
\def\m{\mu}
\def\n{\nu}
\def\o{\over}
\def\om{\omega}
\def\O{\Omega}
\def\p{\phi}
\def\pa{\partial}
\def\pr{\prime}
\def\ra{\rightarrow}
\def\rh{\rho}
\def\rg{\rangle}
\def\s{\sigma}
\def\t{\tau}
\def\th{\theta}
\def\ti{\tilde}
\def\wti{\widetilde}
\def\inte{\int dx }
\def\xb{\bar{x}}
\def\yb{\bar{y}}

\def\tr{\mathop{\rm tr}}
\def\Tr{\mathop{\rm Tr}}
\def\partder#1#2{{\partial #1\over\partial #2}}
\def\ds{{\cal D}_s}
\def\wtwo{{\wti W}_2}
\def\lie{{\cal G}}
\def\alie{{\widehat \lie}}
\def\dlie{{\cal G}^{\ast}}
\def\elie{{\widetilde \lie}}
\def\edlie{{\elie}^{\ast}}
\def\hlie{{\cal H}}
\def\wlie{{\widetilde \lie}}

\def\rlx{\relax\leavevmode}
\def\inbar{\vrule height1.5ex width.4pt depth0pt}
\def\IZ{\rlx\hbox{\sf Z\kern-.4em Z}}
\def\IR{\rlx\hbox{\rm I\kern-.18em R}}
\def\IC{\rlx\hbox{\,$\inbar\kern-.3em{\rm C}$}}
\def\one{\hbox{{1}\kern-.25em\hbox{l}}}

\def\PRL#1#2#3{{\sl Phys. Rev. Lett.} {\bf#1} (#2) #3}
\def\NPB#1#2#3{{\sl Nucl. Phys.} {\bf B#1} (#2) #3}
\def\NPBFS#1#2#3#4{{\sl Nucl. Phys.} {\bf B#2} [FS#1] (#3) #4}
\def\CMP#1#2#3{{\sl Commun. Math. Phys.} {\bf #1} (#2) #3}
\def\PRD#1#2#3{{\sl Phys. Rev.} {\bf D#1} (#2) #3}
\def\PLA#1#2#3{{\sl Phys. Lett.} {\bf #1A} (#2) #3}
\def\PLB#1#2#3{{\sl Phys. Lett.} {\bf #1B} (#2) #3}
\def\JMP#1#2#3{{\sl J. Math. Phys.} {\bf #1} (#2) #3}
\def\PTP#1#2#3{{\sl Prog. Theor. Phys.} {\bf #1} (#2) #3}
\def\SPTP#1#2#3{{\sl Suppl. Prog. Theor. Phys.} {\bf #1} (#2) #3}
\def\AoP#1#2#3{{\sl Ann. of Phys.} {\bf #1} (#2) #3}
\def\PNAS#1#2#3{{\sl Proc. Natl. Acad. Sci. USA} {\bf #1} (#2) #3}
\def\RMP#1#2#3{{\sl Rev. Mod. Phys.} {\bf #1} (#2) #3}
\def\PR#1#2#3{{\sl Phys. Reports} {\bf #1} (#2) #3}
\def\AoM#1#2#3{{\sl Ann. of Math.} {\bf #1} (#2) #3}
\def\UMN#1#2#3{{\sl Usp. Mat. Nauk} {\bf #1} (#2) #3}
\def\FAP#1#2#3{{\sl Funkt. Anal. Prilozheniya} {\bf #1} (#2) #3}
\def\FAaIA#1#2#3{{\sl Functional Analysis and Its Application} {\bf #1} (#2)
#3}
\def\BAMS#1#2#3{{\sl Bull. Am. Math. Soc.} {\bf #1} (#2) #3}
\def\TAMS#1#2#3{{\sl Trans. Am. Math. Soc.} {\bf #1} (#2) #3}
\def\InvM#1#2#3{{\sl Invent. Math.} {\bf #1} (#2) #3}
\def\LMP#1#2#3{{\sl Letters in Math. Phys.} {\bf #1} (#2) #3}
\def\IJMPA#1#2#3{{\sl Int. J. Mod. Phys.} {\bf A#1} (#2) #3}
\def\AdM#1#2#3{{\sl Advances in Math.} {\bf #1} (#2) #3}
\def\RMaP#1#2#3{{\sl Reports on Math. Phys.} {\bf #1} (#2) #3}
\def\IJM#1#2#3{{\sl Ill. J. Math.} {\bf #1} (#2) #3}
\def\APP#1#2#3{{\sl Acta Phys. Polon.} {\bf #1} (#2) #3}
\def\TMP#1#2#3{{\sl Theor. Mat. Phys.} {\bf #1} (#2) #3}
\def\JPA#1#2#3{{\sl J. Physics} {\bf A#1} (#2) #3}
\def\JSM#1#2#3{{\sl J. Soviet Math.} {\bf #1} (#2) #3}
\def\MPLA#1#2#3{{\sl Mod. Phys. Lett.} {\bf A#1} (#2) #3}
\def\JETP#1#2#3{{\sl Sov. Phys. JETP} {\bf #1} (#2) #3}
\def\JETPL#1#2#3{{\sl  Sov. Phys. JETP Lett.} {\bf #1} (#2) #3}
\def\PHSA#1#2#3{{\sl Physica} {\bf A#1} (#2) #3}
\def\PHSD#1#2#3{{\sl Physica} {\bf D#1} (#2) #3}
\begin{titlepage}
\vspace*{-2 cm}
\noindent
\begin{flushright}

\end{flushright}

\vskip 1 cm
\begin{center}
{\Large\bf T-duality of axial and vector  dyonic integrable models   } \vglue 1  true cm
{ J.F. Gomes}$^{\dagger}$, E. P. Gueuvoghlanian$^{\dagger}$,
 { G.M. Sotkov}$^{\dagger}$ and { A.H. Zimerman}$^{\dagger}$\\

\vspace{1 cm}

$^{\dagger}${\footnotesize Instituto de F\'\i sica Te\'orica - IFT/UNESP\\
Rua Pamplona 145\\
01405-900, S\~ao Paulo - SP, Brazil}\\
jfg@ift.unesp.br, gueuvogh@ift.unesp.br, sotkov@ift.unesp.br, zimerman@ift.unesp.br\\

\vspace{1 cm}

\end{center}

\normalsize
\vskip 0.2cm

\begin{center}
{\large {\bf ABSTRACT}}\\
\end{center}
\noindent
A general construction of affine  Non Abelian (NA) - Toda models  in terms of axial
and vector gauged
two loop WZNW model is discussed. They represent {\it integrable perturbations} of the
conformal $\sigma$-models (with tachyons included) describing (charged) black hole
type string backgrounds. We study the {\it off-critical} T-duality between
certain families of axial and vector type of integrable models for the case of
affine NA- Toda theories with one global U(1) symmetry. In particular
 we find the  Lie algebraic condition defining  a subclass 
 of {\it T-selfdual} torsionless NA Toda models 
 and  their zero curvature representation.

\noindent

\vglue 1 true cm

\end{titlepage}

\sect{Introduction}
Two dimensional integrable models represent  an important laboratory
for testing new ideas and developing new methods for constructing exact
solutions as well as for the nonperturbative quantization of 4-D non-abelian
gauge theories, gravity and  string theory.  Among the numerous
techniques for constructing 2-D integrable models   and their solutions
\cite{fadeev}, \cite{lez-sav}, the hamiltonian reduction of the WZNW model 
(or equivalently the gauged WZNW ) 
associated to a finite dimensional Lie algebra $\lie $ has provided an universal 
and simple method for deriving the equations of motion
(or action ) of 2-d integrable models.  In particular, the conformal Toda (CT) 
models were constructed by implementing a consistent set of
constraints on the WZNW currents  \cite{ora}.  The method was subsequently extended to
infinite dimensional affine algebras ( denoted here by $\hat \lie $),
 leading to WZNW currents satisfying the so called two loop current algebra \cite{schw}.  
  By  further imposing an infinite number of suitable constraints 
 the conformal affine Toda (CAT) models were constructed \cite{Aratyn}.  
 The power
of such method was demonstrated in constructing (multi) soliton solutions of the
abelian affine Toda models\cite{Aratyn}
 and certain nonsingular nonabelian (NA) affine Toda
models \cite{luis}.

The present paper is devoted to the systematic construction of a new family of 
{\it axial} and {\it vector}  affine 
NA Toda integrable
models  (associated to an affine  Kac-Moody algebra of rank r - $\hat \lie _r $)
 having  one global U(1) symmetry. They represent appropriate
integrable perturbations of the conformal $\sigma$-models (with tachyons and
dilaton included) describing strings on curved backgrounds of black hole
type (see ref.\cite {ginsparg}, \cite{ts}). 
An important feature of these integrable models  is that they
admit U(1)-{\it charged topological} solitons for imaginary coupling constant 
\cite{galen}. With topological $\theta$-term added to their actions, the one
soliton spectrum manifests properties quite similar to the { dyons} of 4-D
Yang-Mills -Higgs model namely, their electric charges get contributions from
the magnetic (topological) one. The conformal limits 
(without the tachyonic terms) of 
axial and vector models in consideration,
are known to be T-dual  to each other, having $O(r,r|Z)$ as T-duality group 
(see for example \cite{giveon}). The
{ natural question} arises whether one can extend the critical (i.e. conformal)
T-duality transformations to the corresponding integrable model and which  is the
{ off-critical } T-duality group.  The problem of non conformal (i.e. off critical ) 
T-duality was first addressed  in \cite{alv} in the
context of the principal chiral $\s$-models.  For more general discussion 
of T-duality as canonical transformations of conformal and
non-conformal 2-d models see \cite{orlando}. 
It turns out that axial and vector models
similarly to their conformal $\sigma$- models counterparts are forming again
T-dual pairs. The critical T-duality group is however broken 
to $O(1,1|Z)$
for the family of dyonic integrable models  under investigation.

As in the conformal case
the subclass of {\it T-selfdual } models is of particular interest. We first
investigate  the (torsionless) T-selfdual {\it conformal}
 $\lie_r$-NA Toda 
models obtained from the conformal $\sigma$-models (associated to finite dimensional Lie algebras $\lie _r$)
 by adding all the possible marginal operators.
As it shown in Sect.4.1, T-seldual conformal NA -Toda models exist
only for orthogonal algebras $\lie_r = B_r = SO(2r+1)$ and for specific choice of the form of the  marginal
operators  only.
The axial and vector dyonic models  can be obtained by
adding certain integrable relevant operators to the conformal $\lie_r$-NA Toda
actions. 
 As is well known \cite{lez-sav},\cite{luis} the algebraic structure
underlying these { nonconformal} affine $\hat\lie _r$-NA Toda models is 
encodded in the
choice of grading operator Q for $\hat\lie _r$  and in the form of 
the grade $\pm 1$ 
constant elements $\epsilon_{\pm}$ (which characterizes the  potential).
Addressing once more the question of the T-selfdual (torsionless) members
of the considered family of integrable models, we derive the Lie algebraic condition  
(i.e. all the possible choices of the algebraic
 data - $\hat\lie_r, Q,\epsilon_{\pm})$
  that gives rise to T-selfdual  models.
 It shown in sect. 4.2 and 4.3 that
such subclass of models exist only for the following 
{ three} affine Kac-Moody algebras:$ B_r^{(1)},
 A_{2r}^{(2)} $ and $ D_{r+1}^{(2)}$ when  the grading operator Q
  and  the $\epsilon_{\pm}$'s are appropriately choosen.

An interesting byproduct
of our { algebraic T-selfduality } condition is that the above three families of
 torsionless models
exactly reproduce the Fateev's integrable models \cite{Fat}.
 Our construction also provides a simple
and systematic proof of their classical integrability. 
It is worthwhile 
to mention the following simple
form of our T-selfduality condition: while the generic axial and vector models ( with one global U(1)
symmetry) are characterized by the fact that their physical fields 
$g_0^f$
lie in the coset $\lie_0 /{\lie_0^0} ={{SL(2) \otimes U(1)^{rank \lie -1}}\o {U(1)}}$, 
the corresponding cosets for the T-selfdual models are of the
particular form $ \lie_0 /{\lie_0^0} = {{SL(2)}\o {U(1)}} \otimes 
U(1)^{rank \lie -1}$ .

This paper is organized as follows. Sect.2 contains the functional integral
derivation of the effective actions for generic { conformal,affine and
conformal affine} NA -Toda theories. In the particular case when these models
manifest $\lie_0^0 $=U(1) gauge symmetry, the actions for the corresponding 
{ singular} affine NA Toda models of { axial} and { vector} type are 
 obtained. Two particular examples based on the affine algebras $A_r^{(1)}$  and
 $B_r^{(1)}$  are presented. Sect.3 is devoted to the analysis of the abelian {\it
 off-critical} T-duality relating the axial and vector type of models. We derive
 the Lie algebraic condition defining the family of {\it T-selfdual
 torsionless} singular affine NA Toda theories in Sect.4.
 In Sect.5 we present the zero curvature representations of all IM's
 in consideration.

\sect{Gauged WZNW Construction of NA Toda Models}

The generic conformal (or affine) $G_r$ (or $\hat{G}_r$) -  NA - Toda models  are classified
 according to a $\lie_0 \subset \lie$ embedding  induced
by the grading operator $Q$ \cite{lez-sav}, which defines a specific decomposition of the 
corresponding finite $\lie_r$ or  infinite( $\hat{\lie}_r$) dimensional 
Lie algebras 
$\lie = \oplus _{i} \lie _i $ where $
[Q,\lie_{i}]=i\lie_{i}$ and $ [\lie_{i},\lie_{j}]\subset \lie_{i+j}$. The group
element $g$ can then be written in terms of the Gauss decomposition as 
\be
g= NBM
\label{1}
\ee
where $N=\exp \lie_< \in H_{-}$, $B=\exp \lie_{0} $ and
$M=\exp \lie_> \in H_{+}$.  The {\it physical fields } $B$ 
lie in the zero grade subgroup $\lie_0$ 
and the
models we seek correspond to the coset $H_- \backslash G/H_+ $.

For consistency with the Hamiltonian reduction formalism, the phase space of
the G-invariant WZNW model is  reduced by specifying the constant
generators $\eps_{\pm}$ of grade $\pm 1$.  In order to derive 
 an action for $B \in \lie_0 $,  invariant under 
\begin{eqnarray}
g\longrightarrow g^{\prime}=\alpha_{-}g\alpha_{+},
\label{2}
\end{eqnarray}
where $\a_{\pm}(z, \bar z) \in H_{\pm}$ 
 we have to introduce a set of  auxiliary
gauge fields $A \in \lie _{<} $ and $\bar A \in \lie _{>}$ transforming as 
\begin{eqnarray}
A\longrightarrow A^{\prime}=\alpha_{-}A\alpha_{-}^{-1}
+\alpha_{-}\partial \alpha_{-}^{-1},
\quad \quad 
\bar{A}\longrightarrow \bar{A}^{\prime}=\alpha_{+}^{-1}\bar{A}\alpha_{+}
+\bar{\partial}\alpha_{+}^{-1}\alpha_{+}.
\label{3}
\end{eqnarray}
The result is given by the gauged WZNW action (see for instance \cite{gaw}, \cite{ora}),
\br
S_{G/H}(g,A,\bar{A})&=&S_{WZNW}(g)
\nonumber
\\
&-&\frac{k}{2\pi}\int dz^2 Tr\( A(\bar{\partial}gg^{-1}-\epsilon_{+})
+\bar{A}(g^{-1}\partial g-\epsilon_{-})+Ag\bar{A}g^{-1}\) .
\nonu
\er
Since the action $S_{G/H}$ is $H$-invariant,
 we may choose $\alpha_{-}=N^{-1}$
and $\alpha_{+}=M^{-1}$.  From the orthogonality  of the graded 
subspaces, i.e. $Tr( \lie _i\lie _j ) =0, i+j \neq 0$, we find 
\begin{eqnarray}
S_{G/H}(g,A,\bar{A})&=&S_{G/H}(B,A^{\prime},\bar{A}^{\prime})
\nonumber
\\
&=&S_{WZNW}(B)-\frac{k}{2\pi}
\int dz^2 Tr[A^{\prime}\epsilon_{+}+\bar{A}^{\prime}\epsilon_{-}
+A^{\prime}B\bar{A}^{\prime}B^{-1}],
\label{14}
\end{eqnarray}
where 
\begin{eqnarray}
S_{WZNW}=- \frac{k}{4\pi }\int d^2zTr(g^{-1}\partial gg^{-1}\bar{\partial }g)
-\frac{k}{24\pi }\int_{D}\epsilon_{ijk}
Tr(g^{-1}\partial_{i}gg^{-1}\partial_{j}gg^{-1}\partial_{k}g),
\label{3a}
\end{eqnarray}
and the topological term denotes a surface integral  over a ball $D$
identified as  space-time.

The action (\ref{14}) describes  nonsingular Toda models among which we find the
conformal and the affine abelian Toda models for 
$Q=\sum_{i=1}^{r}\frac{2\lambda_{i}\cdot H}{\alpha_{i}^{2}}, \quad 
 \epsilon_{\pm}=\sum_{i=1}^{r}c_{\pm i}E_{\pm \alpha_{i}}$ and 
$Q= h\hat {d}+ \sum_{i=1}^{r}\frac{2\lambda_{i}\cdot H}{\alpha_{i}^{2}}, \quad 
\hat\epsilon_{\pm}=\sum_{i=1}^{r}c_{\pm i}E_{\pm \alpha_{i}}^{(0)} + 
E_{\pm \a_0 }^{(\pm 1)}$
respectively.\footnote { by $-\a_0 $ we denote the highest root, $\lambda_i$ -  the 
fundamental weights, $h$ - the coxeter number of $\lie $
  and $H_i$  are the Cartan subalgebra generators 
in the Cartan - Weyl basis satisfying $ Tr (H_iH_j) = \d_{ij}$.} In both cases the 
zero grade subgroup $\lie_0=U(1)^r$  is abelian
 and it coincides with the Cartan subalgebra of $\lie_r$.
Performing the integration  over the auxiliary fields $A$ and $\bar A$ in 
the functional integral 
\be
 Z_{\pm}=\int DAD\bar{A}\exp (-F_{\pm}),
\label{fpm}
\ee 
 where 
 \be
F_{\pm} = {-{k\o {2\pi}}}\int \(Tr
 (A - B {\eps_-} B^{-1})B(\bar A - B^{-1} {\eps_+} B) B^{-1}\)d^2z
\nonu
\ee
we derive the effective action for the {\it conformal abelian} Toda theories
\be
S = S_{WZNW} (B) - {{k\o {2\pi}}} \int Tr \( \eps_+ B  \eps_- B^{-1}\)d^2z
\label{spm}
\ee
and for the (nonconformal) {\it affine abelian} Toda IM by replacing $\epsilon_{\pm}$ from 
the finite algebra with the $\hat\epsilon_{\pm }$ belonging to
the affine Kac - Moody algebra $\hat \lie$. 
By construction the  later (affine Toda) action  describes  an integrable perturbation of
 the $\lie$-conformal abelian Toda model. 

 More interesting conformal (and  affine) Toda models  
arise when the grading structure (defined by the operator Q) leads to non abelian zero grade 
subalgebras $\lie_0\subset\lie $. In particular, if we 
supress one
of the fundamental weights from $Q$, the zero grade subspace acquires 
a nonabelian structure
  $ sl (2)\otimes u(1)^{rank \lie -1}$.  Let us consider for instance 
$Q= h^{\pr} \hat {d} + \sum_{i\neq a}^{r}\frac{2\lambda_{i}\cdot H}{\alpha_{i}^{2}}$, 
where $h^{\pr} =0$ or
$h^{\pr} \neq 0$ corresponding to
 the conformal or affine  nonabelian (NA) - Toda models
respectively.
The absence of $\lambda_a$ in $Q$
 prevents the contribution of the simple root step operators
$E_{\pm\a_a}^{(0)}$ in constructing $\eps_{\pm}$(or in $\hat\epsilon_{\pm}$), since the generators $E_{\pm\a_a}^{(0)}$ now 
belong to the zero grade
$\lie_0$. The form of the corresponding actions is as (\ref{spm}), but with $\eps_{\pm}$ and $B\in sl(2)\otimes u(1)^{rank \lie -1}$ as described
above. They are known under the name {\it nonsingular} conformal (and affine)  NA - Toda models. An important feature of these models is that 
they manifest an  {\it additional chiral (left and right) } $U(1)$ - symmetry. The  algebraic origin of this fact is in the specific graded
structure that allows the existence of U(1) - generator $\lie_0^0 = Y\cdot H \in \hat \lie_0$  such that $[Y\cdot H , \eps_{\pm}] = 0$.
 The first term in the
action (\ref{spm}) is invariant under local (chiral)  $\lie _0^0$ - transformations by construction, while the invariance of 
the  second (potential) term is a consequence of the defining property of $Y\cdot H$ (i.e. to commute with the $\eps_{\pm}$). 
The  corresponding  (chiral) conserved currents  $ J_{Y \cdot H} $  and $\bar J_{Y \cdot H} $ have 
the standard form $J_{Y\cdot H} = Tr[(Y\cdot H )J]$, and 
$J = g^{-1}\partial g$ and $ \bar{J} = -\bar{\partial}gg^{-1}$. By gauge fixing of this symmetry  one obtains {\it singular} conformal (and affine )
NA - Toda models with the number of physical fields reduced by one. The elimination of the $\lie _0^0$ field  R in the
framework of the Hamiltonian reduction consists in imposing of the {\it nonlocal } constraints $J_{Y \cdot H} = \bar J_{Y \cdot H} = 0$. 
The 
standard method \cite{gaw} of incorporating these constraints in the functional integral (\ref{fpm}) is to introduce auxiliary gauge fields 
$A_{0}=(Y.H)a_0$ and $\bar{A}_{0}=(Y.H)\bar a_0$ in the action (\ref{spm}) in such a way that the constraints
 appear as equations of motion of the 
 improved action. As is well known \cite{gaw} the improvement consists in adding new   
 $A_{0}$, $\bar{A}_{0}$- dependent terms in the manner that the new action to
  be invariant under the following  $\lie_0^0$ -transformations
\br
g\longrightarrow g^{\pr}=\alpha_{0}g\alpha_{0}^{\pr},\quad  
A_{0}\longrightarrow A_{0}^{\pr}=A_{0}-\alpha_{0}^{-1}\partial \alpha_{0},
\quad \quad 
\bar{A}_{0}\longrightarrow \bar{A}_{0}^{\prime}=\bar{A}_{0}
- \bar{\partial}\alpha_{0}^{\pr}(\alpha_{0}^{\pr})^{-1}
\label{5}
\er
There exist two inequivalent cases of gauge fixing of $\lie_0^0 = U(1)$ - symmetry, namely the {\it axial gauging} where  $\a^{\pr}_{0}
 =\alpha_{0}(z, \bar z) \in \lie_0^0 $ and the {\it vector gauging}, where 
$ \a^{\pr}_{0}
 ={\alpha_{0}}^{-1}(z, \bar z) \in \lie_0^0 $. Finaly, the improved action with all these properties has the form \cite{gaw},\cite{ginsparg} :
\br 
S(B,{A}_{0},\bar{A}_{0} ) &=& S(g_0^f,{A}_{0},\bar{A}_{0} )  
 = S_{WZNW}(B)\nonu \\   
  &-&{{k\o {2\pi}}}\int Tr\(\pm  A_{0}\bar{\partial}B
B^{-1} + \bar{A}_{0}B^{-1}\partial B
\pm A_{0}B\bar{A}_{0}B^{-1} + A_{0}\bar{A}_{0} \)d^2z\nonu \\  
&-& {{k\o {2\pi}}} \int d^2z Tr \hat \eps_+ B \hat \eps_- B^{-1} 
\label{aa}
\er
where the $\pm $ signs in (\ref{aa}) correspond to axial/vector gaugings respectively. For generic affine Kac - Moody algebra $\lie _r$, the
zero grade element $B\in
\lie_0$ is parametrized as follow \footnote{ All algebraic notations are as in ref.\cite{ime}. The generators $H_i, E_{\a_a}$,
 when written without upper
indices correspond to $H_i^{(0)}, E_{\a_a}^{(0)}$ respectively. The extra generators $\hat c$ and $\hat d$ in the affine case with 
the properties $[{\hat d} , E^n_{\a_a}] = n E^n_{\a_a },\quad [{\hat d} , H^n_{\a_a}] = n H^n_{\a_a},\quad 
[{\hat c}, E^n_{\a_a}] = [{\hat c},{\hat d} ] = 0 $ represent the center and the derivative respectively.}
\begin{eqnarray}
B=\exp (\tilde {\chi} E_{-\alpha_{a}})
 \exp (   R {{Y^j}} {{H_j}}+\Phi (H)+ \nu \hat {c} +
  \eta \hat {d})\exp (\tilde {\psi} E_{\alpha_{a}})
\label{632}
\end{eqnarray}
where $ \Phi (H) =\sum_{j=1}^{r}\sum_{i=1}^{r-1}\varphi_{i}{{X}^j}_i {{h_j}}$
,  $Y\cdot X_i =\sum_{j=1}^r{{Y^j}}  {{X}^j}_{i} =0, \;\; i=1, \cdots ,r-1$ 
and $ h_j = {{2 \a_j\cdot H} \o {\a_j^2}}, 
 j=1, \cdots ,r$, i.e. $X_i$ denotes r-dimensional vectors  ortogonal to 
the particularly choosen vector Y.
We next define the partition function of the
reduced $\lie_0 /{\lie_0^0} = ({{sl(2)\otimes
u(1)^{rank \lie -1}})/u(1)}$ -  NA - Toda models : 
\be
Z_{sing} = \int DBDA_{0}D\bar{A}_{0}e^{-S(B,{A}_{0},\bar {A}_{0})}
\label{paf}
\ee
Integrating out the auxiliary fields $A_0$, $\bar A_0$ in (\ref{paf}) we derive  the effective action $S_{eff}(g_0^f)$, 
where $g_0^f\in \lie_0 /{\lie_0^0}$ is parametrized by the physical fields of the axial or vector {\it singular} conformal (or affine) 
NA - Toda models. The explicite form of the corresponding effective actions depends on the specific  algebraic data - $\lie_r
,Q,\eps_{\pm}$, i.e. on the algebra (finite or infinite) $\lie_r$ of rank r, on the  root $\a_a $ that is missed in Q, on the choice
of $Y\cdot H = \lie_0^0$ (fixed by Q and the form of $\eps_{\pm}$) and finaly on the way (axial or vector) the $\lie_0^0 (= U(1))$ is gauge
fixed. An important property  common for  the entire family of  conformal and affine singular (axial or vector) NA - Toda models
 with $\lie_0^0 = U(1)$ 
is their {\it global} U(1) Noether symmetry.\footnote{ remember that in the case $\lie _0$ is abelian an invariant subalgebra $\lie_0^0$ does
not exist and the corresponding  affine {\it abelian} Toda models do not have any Noether symmetries.} Examples of singular
 {\it conformal} $\lie_r$ - NA - Toda models of {\it axial type} have been constructed in refs.\cite{ime},\cite{plb} and 
\cite{annals}. In the next two subsections we generalize the conformal constructions of refs.\cite{ime} and 
\cite{annals} to the case of (infinite ) affine algebras as well as for the  case of vector gauging of $\lie _0^0$. The integrable models
 obtained
in this way represent the family of singular affine \footnote{and conformal affine  NA- Toda models when the generators $\hat c$ and $\hat d$ are 
included in the zero
grade subalgebra} NA - Toda models of {\it  axial} and {\it vector} type with one global U(1) symmetry. Their main characteristic appears to be 
the fact \cite{galen} that for imaginary coupling they admit U(1) - charged {\it topological} solitons with the electric and magnetic charges
similar to the dyonic spectrum of certain 4-dimensional Yang-Mills-Higgs theory\cite{galen}. This is the reason to call them {\it dyonic
 integrable models }. As we shall demonstrate in Sect.3 below the fact that the axial and vector IMs are obtained from the unique nonsingular IM
 with local U(1) symmetry  by two different gauge fixings of this symmetry, gives rise to an interesting phenomena : pairs of integrable 
 models related by the {\it off-critical} T-duality transformation.

 \subsection{Axial Gauging}
 
Taking into account the fact that the action (\ref{aa}) is invariant under U(1) transformations (\ref{5}) we can gauge away the nonlocal 
field $R$ 
 by choosing $\a_0^{\pr} =\a_0 = e^{-{1\o 2}Y \cdot H R}$, that corresponds to {\it axial} gauge fixing. Then the gauge fixed element B (i.e.
  the factor group element $g_0^f\in \lie_0 / {\lie_0^0}$) becomes    
\be
g_0^f=\exp (\chi E_{-\alpha_{a}})
 \exp (   \Phi (H)+ \nu \hat {c} + \eta \hat {d})\exp (\psi E_{\alpha_{a}})
 \label{63a}
\ee
where $\chi = \tilde {\chi}e^{{1\o 2}Y\cdot \a_a R}$ and 
$\psi = \tilde {\psi}e^{{1\o 2}Y\cdot \a_a R}$.
With this parametrization the  second ($A_0, \bar{A}_{0}$) - dependent term in the action (\ref{aa}) takes the form :
\br
F_0&=& -\frac{k}{2\pi}\int  Tr\( A_{0}\bar{\partial}g_{0}^{f}
({g_{0}^{f}})^{-1}
+\bar{A}_{0}({g_{0}^{f}})^{-1}\partial g_{0}^{f}
+A_{0}g_{0}^{f}\bar{A}_{0}({g_{0}^{f}})^{-1}+A_{0}\bar{A}_{0} \)d^2z
\nonu  \\
&=&{-{k\o {2\pi}}}\int \( a_0 \bar a_0 2Y^2\Delta_a -  ({{2\a_a \cdot Y}\o
{\a_a^2}})(\bar a_0\psi \pa \chi + a_0 \chi \bar \pa \psi )e^{\Phi
(\a_a)}\)d^2z
\label{del}
\er
where $\Delta_a = 1 + {{(Y \cdot \a_a )^2}\o {{\a_a^2}{Y^2}}}\psi \chi e^{\Phi (\a_a )}$ and $[\Phi (H), E_{\a_a}] =
\Phi(\a_a)E_{\a_a} $.  
As we have mentioned the effective action is obtained by integrating over the auxiliary fields $A_0$ and  $\bar A_0$ in the functional
integral (\ref{paf})
\be
Z_{sing} = \int Dg_0^fDA_{0}D\bar{A}_{0}\exp (F_{0}) \sim \int Dg_0^f e^{-S_0 - S_{WZNW}(g_0^f) - S_{int}}
\label{z0}
\ee
where $S_0 = -{k \o {2\pi}}({{2Y \cdot \a_a }\o {\a_a^2}})^2\int d^2z{{\psi \chi \bar
 \pa \psi \pa \chi }\o
{2Y^2 \Delta_a }}e^{2\Phi(\a_a)}$ and $S_{int} =  {{k\o {2\pi}}} \int d^2z Tr [\hat \eps_+ g_0^f \hat \eps_-( g_0^f)^{-1}]$. The total
 effective action (\ref{aa}) for the axial IM is therefore given as
\br
 S_{eff}&=&-{k \o {4\pi}}\int d^2z\Bigg (Tr(\pa \Phi(H)\bar \pa \Phi(H)) + 
 {{4\bar \pa \psi \pa \chi e^{\Phi(\a_a)}}\o {\a_a^2 \Delta_a}} \nonu\\
&+&\pa \eta \bar \pa \nu +
 \pa \nu \bar \pa \eta 
   - 2 Tr (\hat {\eps_+} g_0^f\hat {\eps_-} (g_0^f)^{-1}) \Bigg )
\label{action}
\er
Note that the second term in (\ref{action}) contains both symmetric and
antisymmetric parts:
\br
\frac{e^{\Phi(\a_a)} } {\Delta_a}\bar{\partial}\psi \partial \chi
=\frac{ e^{\Phi(\a_a)} }
{\Delta_a}(g^{\mu \nu}\partial_{\mu}\psi\partial_{\nu}\chi
+\epsilon_{\mu \nu}\partial_{\mu}\psi \partial_{\nu}\chi )
\nonu
\er
where $g_{\mu \nu}$ is the 2-D metric of signature $ g_{\mu
\nu}= diag (1,-1)$, $\pa = \pa_0 + \pa_1\;\; \bar \pa = \pa_0 - \pa_1$.
 For $r=1$ ($\lie \equiv A_{1}^{(1)}$, $\Phi (\a_1)$ is zero) the
antisymmetric term is a total derivative:
\begin{eqnarray}
\epsilon_{\mu \nu}\frac{\partial_{\mu}\psi \partial_{\nu}\chi}{1+\psi \chi}
=\frac{1}{2}\epsilon_{\mu \nu}\partial_{\mu}
\left( \ln \left\{ 1+\psi \chi \right\}
\partial_{\nu}\ln{\frac{\chi}{\psi}}\right),
\nonu
\end{eqnarray}
and it can be neglected.  This $A_{1}$-NA-Toda model (in the conformal case), 
is known to describe the
2-D black hole solution for (2-D) string theory \cite{Witten1}.
The
$G $-NA conformal Toda models can be used in the
description of specific (r+1)-dimensional black string theories 
\cite{gervais-saveliev}, \cite{ginsparg}
 with  (r-1)-flat and
2-nonflat directions ($g^{\mu
\nu}G_{ab}(X)\partial_{\mu}X^{a}\partial_{\nu}X^{b}$, $X^{a}=(\psi ,\chi
,\varphi_{i})$), containing axions ($\epsilon_{\mu
\nu}B_{ab}(X)\partial_{\mu}X^{a}\partial_{\nu}X^{b}$) and tachyons
($\exp \left\{ -k_{ij}\varphi_{j}\right\} $), as well.
 One particular example of dyonic axial IM based on $A_r^{(1)}$ for $ a = 1$, i.e. $Y\cdot H = \lambda_{1}\cdot H$ will be discussed in Sect.3.

\subsection{Vector Gauging}

The vector gauging is implemented by choosing 
$\a_0^{\pr} =\a_0^{-1} = e^{{1\o 2}   Y \cdot H R^{\pr}}$, where 
$\bar \chi=-e^{{1\o 2}Y \cdot \a_a R^{\pr}}{\tilde \chi }=e^{-{1\o 2}Y \cdot \a_a R^{\pr}}{\tilde \psi }$ ,i.e. 
$R^{\pr}={1\o {Y\cdot \a_a}}ln \Big ({{-\tilde{\psi}} \o {\tilde{\chi}}}\Big )$. Then the factor group element can be  parametrized as
\begin{eqnarray}
g_0^{f}=\exp  ({-\bar \chi} E_{-\alpha_{a}})
 \exp (   \Phi (H)+ \nu \hat {c} +
  \eta \hat {d})\exp  ({\bar \chi} E_{\alpha_{a}})
 \label{631}
 \end{eqnarray}
where $\Phi (H) = Y\cdot H R + \sum_{j=1}^{r-1} \varphi_j X_j^ih_i$.
The second term of the action (\ref{aa}) then takes the form :
\begin{eqnarray}
F_0 &=&{-{k\o {2\pi}}}\int \Bigg ( -{{2 \o {\a_a^2}}}(Y\cdot \a_a)^2 a_0 \bar a_0 
{\bar \chi}^2 e^{\Phi(\a_a)} 
-  \bar a_0 (Y^2\pa R +
{{2Y\cdot \a_a}\o {\a_a^2}}\bar  \chi \pa \bar \chi e^{\Phi(\a_a)}) \nonu \\
&+ & a_0 (Y^2 \bar \pa R + 
{{2Y\cdot \a_a}\o {\a_a^2}}\bar \chi \bar \pa \bar \chi e^{\Phi(\a_a)}\Bigg )d^2z 
\label{delvector}
\end{eqnarray}
Integrating over $a_0$ and $\bar a_0$ in (\ref{paf}) we derive the total effective action 
(\ref{aa}) for the vector gauged IM :
 \begin{eqnarray}
 S_{eff}&=& -{k \o {4\pi}}\int
 \Bigg (\sum_{i,j=1}^{r-1}Tr(X_i\cdot h X_j\cdot h)\pa \varphi_i \bar \pa \varphi_j 
 + Y^2 \pa R \bar \pa R +\pa \eta \bar \pa \nu 
 + \pa \nu \bar \pa \eta \nonu \\
  &+& 2{{2\o {\a_a^2}}}\rho ^2(\a_a){{\pa R \bar \pa R }\o {\bar \chi^2}}e^{-{\Phi(\a_a)}}
 + 2{{2\o {\a_a^2}}}\rho(\a_a)(\pa R \bar \pa  ln \bar \chi + 
  \bar \pa R  \pa  ln \bar \chi)
    - 2 Tr (\hat {\eps_+} g_0^f\hat {\eps_-} (g_0^f)^{-1}) \Bigg )d^2z \nonu \\
 \label{actionvec}
 \end{eqnarray}
where $ \rho (\a_a) = {{Y^2 \a_a^2} \o {2(Y\cdot \a_a)}}$.  
Defining the new variables 
\be
E = e^{\g R}, \quad  F = E^{-1}( 1 -  \d \bar \chi^2 
\exp{\Phi(\a_a)})
\ee
the  action (\ref{actionvec}) becomes
\br
 S_{eff}&= &-{k \o {2\pi}}\int \Bigg ({1\o 2}\sum_{i,j=1}^{r-1}Tr(X_i\cdot h X_j\cdot h)
 \pa \varphi_i \bar \pa \varphi_j + 
 {1\o {2\g^2 }}\pa ln E \bar \pa ln E (Y^2 + 4\g ^2 \Gamma) 
 \nonu \\
 &-& { \Gamma } (\pa ln E \bar \pa \Phi(\a_a) + 
 \bar \pa ln E  \pa \Phi(\a_a) ) 
 + {1\o 2}\pa \eta \bar \pa \nu + {1\o 2}\pa \nu \bar \pa \eta 
 -{\Gamma} {(\pa  E \bar \pa F + \bar \pa E  \pa F)\o {1- EF}}\nonu \\ 
 &-&  Tr (\hat {\eps_+} g_0^f\hat {\eps_-} (g_0^f)^{-1})\Bigg )d^2z
 \label{torsionless}
 \er
 where $2\g  \Gamma  = {2\o {\a_a^2}}\rho (\a_a)$ and $2\d  \Gamma  = {2\o {
 \a_a^2}}$ are chosen in order to eliminate the variable $\bar \chi $.
Notice that the $E,F$- term in the action (\ref{torsionless}) 
is symmetric and {\it contrary} to the axial model the above action is {\it CPT - invariant}. The vector gauging, therefore
provides a construction of torsionless actions T-dual to its axionic 
 counterpart as we shall demonstrate in Sect. 3. 
  This fact raises the question
whether exist T-selfdual torsionless actions,
 i.e.when the axial and vector gauging leads to same action. It is worthwhile to mention that the dyonic integrable models of
  vector type (\ref{torsionless}) represent integrable perturbations of the conformal $\sigma $ - models studied in \cite{ginsparg}
 in the context of the string backgrounds of black hole type . 
 
 \subsubsection{Example1. Torsionless $B_r^{(1)}$ model}
 We consider the  particular case  based on $B_r^{(1)}$ by taking $Y = {{2\lambda_r}\o {\a_r^2}} - {{2\lambda_{r-1}}\o {\a_{r-1}^2}} = e_r,
 \; a=r, \; \rho (\a_r) = {1\o 2}$, 
 where 
 \be
 \lambda_r = {1\o 2} (e_1 + e_2 + \cdots + e_r ), \quad 
 \lambda_{r-1}  =  (e_1 + e_2 + \cdots + e_{r-1} )
 \label{lambda}
 \ee
 and 
 \br
 \eps_{\pm} = \sum_{i=1}^{r-2}E_{\pm \a_{i}}^{(0)} + E_{\pm (\a_{r-1}+ \a_{r})}^{(0)} +
 E_{\mp (\a_1 + 2(\a_2 + \cdots \a_{r-1}+\a_r))}^{(\pm 1)}
\nonu
\er
 Parametrizing the Cartan part of $g_0^f$ as 
  $\Phi(H) = \sum_{i=1}^{r-1} \varphi_i e_i\cdot H +  2(e_r \cdot H) lnE $, we find 
 \br
 S_{eff}&= &-{k \o {2\pi}}\int \Bigg ({1\o 2}\sum_{i,j=1}^{r-1} 
 \pa \varphi_i \bar \pa \varphi_i  
 + {1\o 2}\pa \eta \bar \pa \nu + {1\o 2}\pa \nu \bar \pa \eta 
 - {(\pa E \bar \pa F + \bar \pa E  \pa F)\o {1- EF}}\nonu \\ 
& -&  Tr (\hat {\eps_+} g_0^f\hat {\eps_-} (g_0^f)^{-1})\Bigg )d^2z
 \label{torsionlessBr}
 \er 
 where we  have chosen $\Gamma = 1, \;\; \g= {1\o 2}, \;\; \d =1$. When the pair of fields $\eta$ and $\nu$  are taken to  be equal to zero the
 corresponding conformal affine $B_r^{(1)}$ NA - Toda  model takes the form of the {\it nonconformal} affine $B_r^{(1)}$ NA - Toda  model. 
 
 \subsubsection{Example2.  $A_r^{(1)}$ vector model}
In order to find the explicit form of the singular affine NA - Toda IM based on the affine algebra $A_r^{(1)}$ we take $Y = \lambda_1$ and $\eps _{\pm} = \sum_{i=2}^{r} E_{\pm \a_i}^{(0)} +
  E_{\mp (\a_2 + \cdots \a_r)}^{(\pm 1)}$.
   We will get simpler result if we parametrize 
   $\Phi(H) = \sum_{i=1}^{r} \phi_i h_i$ :
  \br
 S_{eff}&=& -{k \o {4\pi}}\int
 \Bigg (\sum_{i,j=1}^{r}\eta_{ij}\pa \phi_i \bar \pa \phi_j 
 + 2 {{\pa \phi_1 \bar \pa \phi_1}\o {\bar {\chi }^2e^{2\phi_1-\phi_2}}} \nonu \\
  &+& 2(\pa \phi_1 \bar \pa  ln \bar \chi + 
  \bar \pa \phi_1  \pa  ln \bar \chi)
    - 2 Tr (\hat {\eps_+} g_0^f\hat {\eps_-} (g_0^f)^{-1}) \Bigg )d^2z
 \label{actionvec1}
 \er
 With the following change of the field variables
 \be
 E = e^{\phi_1} , \quad F = E^{-1}( 1 -  \bar \chi^2 e^{2\phi_1-\phi_2})
 \ee
 we  find
 \br
 S_{eff} = -{k \o {2\pi}}\int \Bigg ({1\o 2}\sum_{i,j=2}^{r}\eta_{ij}
 \pa \phi_i \bar \pa \phi_j 
 -{1\o 2}{(\pa E \bar \pa F + \bar \pa E  \pa F)\o {1- EF}}
 -  Tr (\hat {\eps_+} g_0^f\hat {\eps_-} (g_0^f)^{-1})\Bigg )d^2z
 \label{Ar}
 \er
 Introducing  new fields $ c_i = e^{\phi_{i+2} -\phi_{i+1}}, \;\; i=1, \cdots r-1$ we 
 rewrite the action (\ref{Ar}) in the form
 \br
 S_{eff}&=& -{k \o {2\pi}}\int \Bigg ({1\o 2}\sum_{i=1}^{r-1}
 \( \pa ln c_i  \bar \pa ln c_1 c_{i+1} \cdots  c_{r-1} + 
 \bar \pa ln c_i   \pa ln c_1 c_{i+1} \cdots  c_{r-1}\) 
 \nonu \\
 &-& {1\o 2}{(\pa E \bar \pa F + \bar \pa E  \pa F)\o {1- EF}}
 -  Tr (\hat {\eps_+} g_0^f\hat {\eps_-} (g_0^f)^{-1})\Bigg )d^2z
 \label{Ar1}
 \er
 These are the vector IM's studied in ref.\cite{galen} .

 \sect{Off-Critical T-Duality}
 
 T-duality\cite {giveon} is known to be an important property of the string
 theory. It acts as canonical transformation \cite {alv-gaume},\cite{giveon} in the
 string phase space ${\cal P} = \{ X^{M}(\s ), \Pi_{M}(\s ) =g_{MN}\dot
 X^{N}+b_{M}^{N}X^{\pr}_{N}; b, c \} $
  mapping the original
 conformal $\sigma $-model \footnote{  under certain symmetry
 restrictions on the geometrical data:$e_{MN}(X)=g_{MN}(X)+b_{MN}(X)$  and 
 $\varphi (X)$. In the
 case of abelian T-duality $e_{MN}(X)$ and  are $\varphi (X)$
 independent of $d \leq D$ of $X_{\a}$, $(\a = 1,2, \cdots d\leq D)$ 
  called isometric target-space
 coordinates.}:
\br
S^{conf}_{\s }= {{1}\o {4\pi \a^{\pr}}}\int d^2z \( (g_{MN}(X)\eta^{\mu \nu} +
b_{MN}(X) \eps^{\mu \nu}) \pa_{\mu}X^{M}\pa_{\nu}X^{N} + {{\a^{\pr}}\o 2}
R^{(2)}\varphi (X) \)
\label{3.1}
\er
($\mu , \nu = 0,1; M,N = 1,2, \cdots D$ and $R^{(2)}$  is the worldsheet 
 curvature ) to its
 T-dual model $\tilde S^{conf}_{\s }(G_{MN}(\tilde X), B_{MN}(\tilde X), \phi
 (\tilde X))$. Curved string backgrounds with
 d-isometric directions provide an example of an  abelian T-duality
 transformation\cite{busher},\cite{giveon}:
 \br
 E_{\a \b }&=&(e^{-1})_{\a \b}, \;\; E_{mn} = e_{mn} - e_{m \a }(e^{-1})^{\a
 \b}e_{\b n} \nonu \\
  E_{\a m }&=&(e^{-1})_{\a}^{ \b}e_{\b m}, \;\; E_{m \a} = - e_{m
  \b}(e^{-1})^{\b}_{\a} \nonu \\
  \phi &=& \varphi - ln [det E_{\a \b}], \;\; \a, \b = 1,2, \cdots d, \;\; m,n
  = d+1, \cdots D
  \label{3.2}
  \er
 where $E_{MN} = G_{MN} + B_{MN}$. The canonical transformation $(\Pi_X, X)
 \rightarrow (\Pi_{\tilde X}, \tilde X)$
  that generates the background maps (\ref{3.2}) has the
 following simple form \cite{alv-gaume}:
 \br
 \Pi_{\tilde X_{\a}} = -  X_{\a}^{\pr}, \quad \quad
  \Pi_{ X_{\a}} = -  \tilde X_{\a}^{\pr}
  \label{3.3}
  \er
  and all the $\Pi_{ X_{m}}$ and $X_m$ remain unchanged. By construction both
 $\sigma $-models $S_\s^{conf}(e, \varphi )$  
 and $\tilde S_\s^{conf}(E, \phi )$ have coinciding energy spectrum and
 partition functions. The corresponding Lagrangeans are related by 
 the generating
function ${\cal F}$:
 \br
 {\cal L}(e,\varphi )&=& {\cal L}(E, \phi) + {{d {\cal F}}\o {dt}} ,\quad
 {\cal F}= {{1\o {8\pi \a^{\pr}}}}\int dx \( X \cdot \tilde X^{\pr}- 
 X^{\pr} \cdot \tilde X \) , 
 \nonu \\
 {{\d {\cal F}}\o {\d X^{\a}}} &=& \Pi_{X_{\a}}, \quad 
 {{\d {\cal F}}\o {\d \tilde X^{\a}}} = -\Pi_{\tilde X_{\a}}
 \label{3.4}
 \er
 An important feature of the abelian T-duality (\ref{3.2}) and (\ref{3.3}) is
 that it maps the $U(1)^{d}$ Noether charges $Q^{\a} = \int_{-\infty}^{\infty}
 J_0^{\a} dx$ of $S_{\s}^{conf}$
  into the topological charges $\tilde Q^{\a}_{top} = 
  \int_{-\infty}^{\infty}dx \pa_{x} \tilde X^{\a}$  of the dual model
 $\tilde S_{\s}^{conf}$:
 \br
 J_{\mu}^{\a} &=& {1\o 2}e^{\a \b} (X_n) \pa _{\mu}X_{\b} + {1\o 2} e^{\a
 m}(X_n)\pa_{\mu} X_{m} \equiv {1\o 2} \eps _{\mu \nu } \pa^{\nu}\tilde X^{\a}
 \nonu \\
 \tilde J_{\mu}^{\a} &=& {1\o 2}E^{\a \b} (\tilde X_n) \pa _{\mu}\tilde X_{\b}
  + {1\o 2} E^{\a
 m}(\tilde X_n)\pa_{\mu} \tilde X_{m} \equiv {1\o 2} \eps _{\mu \nu } 
 \pa^{\nu} X^{\a}
 \label{3.5}
 \er
i.e. $(Q^{\a}, Q^{\a}_{top}) \rightarrow  (\tilde Q^{\a}_{top}, \tilde Q^{\a})$.  
It is well
 known \cite{giveon}, \cite{ginsparg} that the main part of the conformal
 $\sigma $-models representing relevant string backgrounds can be derived
 from the axial or vector gauged $G/H$-WZNW models. 
 All the models constructed
 in Sect.2 with {\it vanishing potential} term $ V= {{m^2k}\o
 {2\pi}}Tr(\eps_+ B \eps_- B^{-1})$    (i.e. $m=0$ )
 are of this type. They have $d=r$ isometric directions, i.e. $e_{mn} $ do not
 depend on $\varphi_i, \;\; (i=2, \cdots ,r)$  and $\theta = {1\o 2}ln {{\psi
 }\o {\chi }}$. The T-duality group in this case is known
 to be $O(r,r|Z)$ (see for example \cite{giveon}).
   Adding the potential $V$ with
 $\eps_{\pm} = \sum_{i \neq a} E_{\pm \a_i}^{(0)}$ (see eqs. (\ref{action})
  and (\ref{torsionless})) specific for the {\it
 conformal} NA -Toda theories with one global U(1) symmetry, we are decreasing
 the number of the isometric coordinates from $d=r$ to $d_0 =1$. 
 Taking   $\eps_{\pm} = \sum_{i \neq a, b} E_{\pm \a_i}^{(0)}$ one
 can construct NA -Toda theories with $d_0=2$, etc.
 
  The { problem } we are addressing in this section is about the T-duality
  between $d_0=1$ axial and vector { integrable (nonconformal)} models
  of Sect.2.1 and Sect.2.2 with potential terms constructed by taking
  \br
  \hat {\eps}_{\pm} &=& \sum_{i =2}^r E_{\pm \a_i}^{(0)} + E_{\mp (\a_2 + \cdots +
  \a_r)}^{(\pm 1)} \nonu \\
  V_a  &=&{{m^2k}\o {2\pi }}\( \sum_{i=2}^{r} e^{\varphi_{i-2}+\varphi_{i} 
  -2\varphi_{i-1}} + e^{\varphi_{2}+\varphi_{r}}(1+ \psi \chi e^{-\varphi_2})\), \;\;
  \varphi_0 = \varphi_{r+1}=0
   \nonu \\
  V_{vec}  &=&{{m^2k}\o {2\pi }}\( \sum_{i=3}^{r} e^{\phi_{i-1}+\phi_{i+1} 
  -2\phi_{i}} +E e^{-2\phi_2 + \phi_3} + F e^{\phi_2 + \phi_{r}} \), 
  \label{3.6}
  \er
 for the $A_r^{(1)}$ model of Example 2.  For the $B_r^{(1)}$ model (with $\eta = 0$)
 of Example 1 (see
  also Sect.4.2) we have 
  \br
  \hat {\eps}_{\pm} = \sum_{i =1}^{r-2} E_{\pm \a_i}^{(0)} + 
  E_{\pm (\a_{r-1} + \a_r)}^{(0)} +
  E_{\pm \a_0 }^{(\pm 1)}, \;\;\;\;  \a_0 + \a_1 + 2(\a_2 + \cdots \a_{r-1}+
  \a_{r})=0
  \label{3.7}
  \er
  with potential given by eq.(\ref{affV})for the axial model. Its 
  vector counterpart is given by
  \br
  V_{vec}^{B_r^1}=\sum_{i=1}^{r-2}|c_i|^2 e^{-\varphi_{i}+\varphi_{i+1}}+
 2|c_{r-1}|^2 (2EF-1)e^{-\varphi_{r-1}}
+|c_{r}|^2e^{\varphi_{1}+\varphi_{2}}
\nonu
\er
 In both cases the axial IM's isometric
 coordinate is $X = \theta = {1\o 2}ln {{\chi} \o {\psi}}$ 
 ($\tilde u^2 = {{\psi} \o {\chi}}$). 
 For the corresponding vector IM's we choose
$\tilde X_{B_r^{(1)}} = R_B = 2 ln E$  and 
$\tilde X_{A_r^{(1)}} = R_A = {{r+1}\o {r}} ln E$ as isometric coordinates.
 It is important to mention
 that in the case of $A_r^{(1)}$ vector model the canonical transformation
 (\ref{3.3})
 with $d=1$ { has to be accompanied} by the following point 
 transformation:
\br
\phi_j = \phi_j^{\pr} - {{r-j+1}\o {2r}}R_A, \quad j=2, \cdots r
\label{3.8}
\er
Then performing  $d_0=1$ T-duality transformation (\ref{3.2}) 
(together with the 
$A_r^{(1)}$ fields transformation (\ref{3.8})) we realize that
${\cal L}_{vec}$ and ${\cal L}_{a}$ given by
eqs. (\ref{torsionless}) and (\ref{action}), with potentials (\ref{3.6})
 and (\ref{affV}),  are related by eq.(\ref{3.4})
with
\br
{{d {\cal F}}\o {dt}} = 2 \Gamma (\pa ln \tilde u \bar \pa ln E -
\bar \pa ln \tilde u \pa ln E )
\nonu 
\er
Notice that the $B_r^{(1)}$ vector and axial Lagrangeans have the same form, i.e. they
are { T-selfdual}.

An alternative way to perform the T-duality transformation between $d_0=1$
axial and vector IM's in consideration consists in making the following
{ nonlocal} change of the field variables:

a)$A_r^{(1)}$ case
\br
E = e^{{{r}\o {r+1}}R_A}, \quad F = e^{-{{r}\o {r+1}}R_A}(1+ \psi \chi
e^{-\varphi_2}), \quad \phi_j = \varphi_j + {{r-j+1}\o r}R_A, \;\; j=2, \cdots r
\label{3.9}
\er

b)$B_r^{(1)}$ case
\br
E = e^{{1 \o 2}R_B}, \quad F = e^{-{1\o 2}R_B}(1+ \psi \chi), 
\quad \phi_j = \varphi_j 
\label{3.10}
\er
instead of the canonical transformation(\ref{3.3})
 ( resulting in (\ref{3.2})). Eqs. (\ref{3.9})
and (\ref{3.10}) in fact represent the integrated form of (\ref{3.3}). 
Their derivation (see
Sect.5 of ref. \cite{galen}) is based on the comparison of the 
$g_0 $ (or $B$) group
elements written in axial and vector parametrizations (\ref{63a}) and
 (\ref{631}), i.e.
imposing $g_0^{vec} = g_0^{ax}$.
 An important ingredient of this calculation are the
relations (\ref{3.5}) between the $U(1)$ currents and the topological
currents $\eps^{\mu \nu }\pa_{\nu}R $ (and $\eps^{\mu \nu }\pa_{\nu}\theta  $).
Note that $R= 2 ln E$  is a nonlocal
(nonphysical) field in the axial model, but it appears to be physical  in the
vector model. In the case of the $B_r^{(1)}$ model (\ref{torsionlessBr}) 
the $U(1)\leftrightarrow $ topological-currents
relations (\ref{3.5}) take the following explicit form:
\br
\pa ln \tilde u &=& {{v \o {v-1}}}{{\pa ln E }} - {1\o 2}{{\pa v }\o {v-1}} \nonu \\
\bar \pa ln \tilde u &=& -{{v \o {v-1}}}{{\bar \pa ln E }} +
  {1\o 2}{{\bar \pa v }\o {v-1}} 
\label{3.11}
\er
where $v = EF$ and $\tilde u^2 = {{\chi }\o {\psi}}$.

Although the T-duality between the vector and axial integrable models is quite
similar to the conformal "free" case (i.e. $V=0$ )\footnote{ the only new
feature is that one should take care about the specific "point"
 transformations
involving the potential $V $ and that the isometric coordinates are 
reduced from $d=r$
to $d_0=1$.} the {\it off-critical T-duality} addresses few new problems
specific for the integrable models. In the case of imaginary coupling constant
 $\b^2 = -{{2\pi }\o k}$, i.e. $\b \rightarrow i\b_0$ and $ \varphi_k
 \rightarrow i \b_0 \varphi_k $, $\psi \rightarrow i \b_0 \psi$,  etc. 
 one expects that both axial and
vector IM's possess {\it soliton} solutions. 
One might wonder what is the
relation between the solitons (and breathers ) of the T-dual integrable
models, whether their soliton spectra coincides (modulo the interchanges
 $Q \rightarrow \tilde Q_{top}, \quad  \tilde Q \rightarrow  Q_{top}$      
        ) and finally about the $O(1,1|Z)$ symmetry of the
solitons energies and massess. Partial answer of all these questions is
presented in our recent work \cite{galen}.
 

 \sect{No Torsion Theorem}

It becomes clear from our discussion in Sect.3 that the abelian T-duality 
between the axial and the
vector dyonic IM's allows to single out two classes of models : (a) T-selfdual torsionless IM's and (b)
pairs of T-duals axial and vector IM's. One may wonder what is the Lie algebraic condition defining the
models of class (a). In answering this question we first establish the T-selfduality condition for the
conformal $d_0=1$ (i.e $\lie_0^0 =U(1) $) singular NA Toda models. We next generalize the conformal
no torsion (T-selfduality)  theorem to the case of U(1) symmetric affine NA Toda IM's of sects. 4.2 and
4.3. 

\subsection{Conformal T-selfdual NA Toda models}

T-selfduality requires that axial and vector models have coinciding 
Hamiltonians. Since the vector
models are by construction torsionless the question to be addressed is about
 the algebraic condition
 under which the axial gauging generate torsionless models as well.

Consider  a finite dimensional Lie algebra $\lie $ with grading operator given by $
Q_{a}=\sum_{i\neq a}^{r}\frac{2}{\alpha_{i}^{2}}{\lambda_i}\cdot H$
and take {\it the most general} constant generators of grade $\pm 1$, i. e.,
\begin{eqnarray}
\epsilon_{\pm}=\sum_{i\neq a}^{r}c_{\pm i}E_{\pm \alpha_{i}}
+b_{\pm}E_{\pm (\alpha_{a}+\alpha_{a+1})}
+d_{\pm}E_{\pm (\alpha_{a}+\alpha_{a-1})}.
\label{60}
\end{eqnarray}
It is clear that if $c_{\pm i},b_{\pm},d_{\pm}\neq 0$, there shall be {\it no}
$\lie_{0}^{0}$ {\it commuting} with $\epsilon_{\pm}$, since  that 
requires an orthogonal direction to all roots appearing in $\epsilon_{\pm}$. 
These are the generalized {\it non-singular } NA-Toda models of ref.
\cite{ora},\cite {luis}.  
The NA-Toda  models of singular metric $G_{ij}(X)$   
correspond to the cases when $\lie_0^0 = U(1)$ ( or $U(1)^r$ in general )
 and $J_{Y \cdot H} 
=0=\bar J_{Y \cdot H}$  imposed
  as a subsidiary
constraint \footnote{If we leave $\lie_{0}^{0}$ unconstrained the resulting
 model belongs again to the  non singular NA-Toda class of models \cite{ora},\cite{luis}. From the string theory
 point of view the class of singular NA Toda models are of great interest since they describe strings
 on specific black hole backgrounds.}.  Depending
 on the choice of the constants $c_{\pm i}, b_{\pm}$ and  $d_{\pm}$ we
 distinguish four families of {\it singular conformal} NA-Toda models:
\vskip.3cm
(i)$b_{\pm}=d_{\pm}=0$, $\lie_{0}^{0}=\frac{2}{\alpha_{a}^{2}}\lambda_a \cdot H$;

(ii)$c_{\pm (a-1)}=c_{\pm (a+1)}=0$,
$\lie_{0}^{0}=\frac{2}{\alpha_{a}^{2}}\lambda_a \cdot H
-\frac{2}{\alpha_{a-1}^{2}}\lambda_{a-1} \cdot H
-\frac{2}{\alpha_{a+1}^{2}}\lambda_{a+1} \cdot H$;

(iii)$c_{\pm (a+1)}=d_{\pm}=0$,
$\lie_{0}^{0}=\frac{2}{\alpha_{a}^{2}}\lambda_a \cdot H
-\frac{2}{\alpha_{a+1}^{2}}\lambda_{a+1} \cdot H$;

(iv)$b_{\pm}=c_{\pm (a-1)}=0$,
$\lie_{0}^{0}=\frac{2}{\alpha_{a}^{2}}\lambda_a \cdot H
-\frac{2}{\alpha_{a-1}^{2}}\lambda_{a-1} \cdot H$.

Of course, if $c_{\pm j} = 0$, $j\neq a, a\pm 1$, we find $\lie_0^0 = 
\lambda_j \cdot  H$.  However, since $[ \lambda_j \cdot H, E_{\pm \alpha_a}] = 0 $,  there
will be no singular metric  present and this case shall be neglected.  
Cases (i) and (ii) are equivalent, since they are related by the Weyl
 reflection
$ \sigma_{\alpha_{a}}(\alpha_{a\pm 1})=\alpha_{a}+\alpha_{a\pm 1}$ and the
corresponding 
 fields are related by non-linear change of the variables. This case has
already been discussed in refs. \cite{plb} and \cite{annals}, and  shown
 to present always the {\it antisymmetric term},
originated by the presence of $e^{k_{ai}\varphi_i}$ in $\Delta_a$  and in the
kinetic term as well.
Since we are removing all dependence in $\lie_{0}^{0}$, when parameterizing
$g_{0}^{f}$, cases (iii) and (iv) may be studied together with
\begin{eqnarray}
g_{0}^{f}=\exp (\chi E_{-\alpha_{a}})
 \exp (   \Phi (H))\exp (\psi E_{\alpha_{a}})
 \label{63}
 \end{eqnarray}
where $ \Phi (H) =\sum_{i=1}^{a-2}\varphi_{i}h_i
+ \varphi_{-}(\chi_{-} \cdot H) + \varphi_{+}(\chi_{+}\cdot H)
+\sum_{i=a+2}^{r}\varphi_{i}h_i $,

\begin{eqnarray}
\chi_{-}^{(iii)}=\alpha_{a-1}+\alpha_{a}, \quad \quad \chi_{+}^{(iii)}=\alpha_{a+1},\quad ,\quad
\chi_{-}^{(iv)}=\alpha_{a-1},\quad \quad \chi_{+}^{(iv)}=\alpha_{a}+\alpha_{a+1}
\label{64}
\end{eqnarray}
for cases $(iii)$ and $(iv)$ respectively, and 
$\lie_{0}^{0}= Y\cdot H$,
such that
$ Tr(\chi_{\pm} \cdot H \lie_{0}^{0})=0 $.
Such parametrization of $g_0^f$ yields 
\begin{eqnarray}
\Phi (\alpha_{a})=\sum_{i=1}^{a-2}k_{ai}\varphi_{i}
+(\alpha_{a}\cdot \chi_{-})\varphi_{-}+(\alpha_{a}\cdot \chi_{+})\varphi_{+}
+\sum_{i=a+2}^{r}k_{ai}\varphi_{i}
\nonu 
\end{eqnarray}
Now, if we consider Lie algebras whose Dynkin diagrams connect only nearest
neighbours, i. e.,
$\Phi (\alpha_{a})=(\alpha_{a}\cdot \chi_{-})\varphi_{-}
+(\alpha_{a}\cdot \chi_{+})\varphi_{+}$,
then the ``no-torsion condition'' implies $\Phi (\alpha_{a})=0$.
Considering case (iii), we have
\begin{eqnarray}
\alpha_{a}\cdot \chi_{-}=\alpha_{a}\cdot (\alpha_{a-1}+\alpha_{a})=0, 
\quad \alpha_{a}\cdot \chi_{+}=\alpha_{a}\cdot (\alpha_{a+1})=0.
\label{74}
\end{eqnarray}
In this case, the only solution for both equations  is to take $a=r$ (in
such a way that $\alpha_{r+1}=0$) and ${\lie}=B_{r}$ (so that
$\alpha_{r}^{2}=- \alpha_{r-1}\cdot \alpha_{r} =1$). This is precisely the case
proposed by Leznov and Saveliev \cite{leznov-saveliev} and subsequently discussed by 
Gervais and Saveliev \cite{gervais-saveliev} and also by Bilal \cite{bilal}, for the  
particular case of $B_{2}$.
For case (iv), the ``{\it no-torsion condition}'' requires that
\br
\alpha_{a-1}\cdot \alpha_{a}=0, \quad \quad 
\alpha_{a}\cdot (\alpha_{a}+\alpha_{a+1})=0,
\nonu 
\er
Both are satisfied for $a=1$ , ${\lie}=C_{2}$, since
$\alpha_{a-1}=0$ and also $\alpha_{1}^{2}= - \alpha_{1}\cdot \alpha_{2}  = 1$,
respectively.

In general, the ``{\it no-torsion condition}'' (T-selfduality), i. e.,
$  \Phi (\alpha_{a})=0  $,
 may be expressed in terms of the structure of the co-set
$\lie_{0}/{\lie}_{0}^{0}=\frac{u(1)^{r-1}\otimes sl(2)}{u(1)}$. The
crucial ingredient for the appearence of $\Phi (\alpha_{a})$ arises from the
conjugation
\br
Tr(A_{0}g_{0}^{f}\bar{A}_{0}({g_{0}^{f}})^{-1}+A_{0}\bar{A}_{0})=2\lambda_{a}^{2}
\left( 1+\frac{2}{\alpha_{a}^{2}}
\frac{\chi \psi \exp (\Phi (\alpha_{a}))}{2\lambda_{a}^{2}}\right).
\nonu 
\er
Henceforth, if all generators belonging to the Cartan subalgebra 
parameterizing $g_{0}^{f}$ commute with $E_{\pm \alpha_{a}}$, then
$\Phi (\alpha_{a})=0$, and therefore the structure of the co-set
\begin{eqnarray}
\frac{{G}_{0}}{{G}_{0}^{0}}=\frac{u(1)^{r-1}\otimes sl(2)}{u(1)}
=u(1)^{r-1}\otimes \frac{sl(2)}{u(1)}
\label{80}
\end{eqnarray}
is the general condition for the {\it absence} of the {\it antisymmetric}
 term in the
action.

Summarizing, 
for finite  Lie algebras, it was shown  that the
{\it absence } of the antisymmetric terms in the action of the axial NA  - Toda models
 can only occur for $\lie  = B_r$, $a=r$ and 
 $\eps_{\pm} = 
\sum_{i=1}^{r-2}c_{\pm i}E_{\pm \alpha_{i}}
+d_{\pm}E_{\pm (\alpha_{r}+\alpha_{r-1})}$.  In such case, $\lie_0^0$ is
generated by $Y\cdot H = (\frac{2\lambda_{r}}{\alpha_{r}^{2}}
-\frac{2\lambda_{r-1}}{\alpha_{r-1}^{2}})\cdot H$ and 
$ \Phi (H) =\sum_{i=1}^{r-2}\varphi_{i}h_i
+  \varphi_{-}(\a_{r-1}+\a_r ) \cdot H $.
Due to the root structure of $B_r$, we verify that $ \Phi (\a_r) = \a_r \cdot
(\a_{r-1} + \a_r)\varphi_{-}=0$.


In order to  extend the {\it no torsion theorem (T-selfduality) } to the case of infinite affine Lie algebras (i.e
for the integrable perturbations of the conformal models of sect. 4.1) let
 we consider
   $h^{\pr} = 1- \sum_{i \neq r} {2\o {\a_i^2}}\lambda_i \cdot \a_0 $ , 
   where $-\a_0 $ is the  highest root of $\lie$ such that
    $\a_0 \cdot (\frac{2\lambda_{r}}{\alpha_{r}^{2}}
-\frac{2\lambda_{r-1}}{\alpha_{r-1}^{2}})=0 $ and   the gradation
$Q_a(h^{\pr})$ that preserves the zero grade subalgebra $\lie_0$, 
(apart from $\hat c$ and $\hat d$).  
 We choose the cooresponding (affine) grade $\pm$ elements in the form 
 $\hat \eps_+ = \eps_+ + E_{\a_0}^{(1)}$.
  Since conformal and the affine models differ only
by the potential term, the solution for the no torsion condition is also
satisfied for infinite dimensional algebras, whose Dynkin diagram possess a
$B_r$-``tail like''.
An obvious solution  is the untwisted $B_r^{(1)}$ model. Two new other  solutions
are given by the twisted affine Kac-Moody algebras $A_{2r}^{(2)}$ and
$D_{r+1}^{(2)}$ as we shall describe in detail in the next two subsections.

\subsection{The $B_r^{(1)}$ Torsionless Affine NA Toda models}

In order to generalize the $ B_r $  conformal models to the affine 
$B_r^{(1)}$ ones we take \\ 
$ Q = 2(r-1) \hat {d} + 
\sum_{i=1}^{r-1}\frac{2\lambda_{i}\cdot H}{\alpha_{i}^{2}}$, which decomposes
$B_r^{(1)}$ into graded subspaces. 
 In particular, the zero grade subspace  turns out  to be $\lie_0 = SL(2) \otimes
U(1)^{r-1}$, generated by
($E_{\pm \a_r}^{(0)}, h_1, \cdots , h_r$). 
Following the arguments of 
the conformal  no torsion theorem of the sect. 4.1 , we have to choose  
\br
 \hat {\eps_{\pm}} = \sum_{i=1}^{r-2} c_{\pm i}E_{\pm \a_i}^{(0)} +
 c_{\pm (r-1)}E_{\pm (\a_{r-1}+\a_r)}^{(0)} + c_{\pm r}E_{\pm \a_0 }^{(\pm 1)} 
\nonu 
\er
 where $ -\a_0$ (defined by $\a_0 + \a_1 + 2(\a_2 + \cdots + \a_{r-1}
 + \a_r)=0$ ) is the highest root of $B_r$ and
 $\lie_0^0$ is
generated by $Y\cdot H = (\frac{2\lambda_{r}}{\alpha_{r}^{2}}
-\frac{2\lambda_{r-1}}{\alpha_{r-1}^{2}})\cdot H$ 
( such that $[{{Y \cdot H}}, \hat {\eps_{\pm}}] = 0$).  The coset $\lie_0
 /\lie_0^0$ (with $\eta=0$) is then parametrized according to (\ref{63a}) with $\Phi(H) =
 \sum_{i=1}^{r-1} {\cal H}_i \varphi_i $,
where ${\cal H}_i = (\a_r + \cdots \a_i )\cdot H$ so that 
$Tr ({\cal H}_i {\cal H}_j) = \d_{ij}, i,j=1, \cdots , r-1$ and 
the total
effective action becomes
\be
S=-\frac{k}{4\pi}\int d^{2}z
\( \frac{1}{4}\sum_{i=1}^{r-1}g^{\mu \nu}\pa_{\mu }\varphi_{i}\pa
_{\nu}\varphi_{i}+
g^{\mu \nu }\frac{\pa _{\mu}\psi \pa _{\nu}\chi}{1+\psi \chi}  - 2V \)
\label{affaction}
\ee
 The ``affine potential'' V for $(n>2)$  has the form 
\be
V=\sum_{i=1}^{r-2}|c_i|^2 e^{-\varphi_{i}+\varphi_{i+1}}+
 2|c_{r-1}|^2(1+2\psi \chi )e^{-\varphi_{r-1}}
+|c_{r}|^2e^{\varphi_{1}+\varphi_{2}}
\label{affV}
\ee
 The case $r=2$, i.e. ${\widehat \lie } =
{\widehat SO}(5)$ has to be considered separately. Choosing  now the grade $\pm 1$ constant elements as
 $ \hat {\eps_{\pm}} = E_{\a_1 +\a_2 }^{(0)} 
+E_{-\a_1 -\a_2 }^{(1)}$, denoting  $\Phi(\a_{r-1}) = \varphi $  and by  further  changing the variables
\br
\psi \longrightarrow i\psi; \quad \chi \longrightarrow i\psi^* ; \quad
\varphi \longrightarrow  i\varphi  \nonu
\er
we derive the following  {\it real} action  
\be
S=-\frac{k}{4\pi}\int d^{2}z
\( {{g^{\mu \nu}\pa _{\mu}\psi \pa _{\nu}\psi^* }\o {(1-\psi \psi^* )}}+
{1\o 4}g^{\mu \nu } \pa _{\mu }\varphi  \pa _{\nu} \varphi 
+ 8(1-2\psi \psi^* )cos \varphi \) 
\label{so5action}
\ee
\subsection{ The twisted affine NA Toda Models}

The twisted affine Kac-Moody algebras are constructed from a finite
dimensional algebra possessing a nontrivial symmetry of their Dynkin
diagrams (this procedure is known as folding).  Such symmetry can be extended to the algebra by 
  an outer automorphism $\sigma $ \cite{cornwell}, 
   as 
\be
\s ( E_{\a}) = \eta_{\a} E_{\s (\a)}
\label{aut}
\ee
where $\eta_\a = \pm 1$.  For the simple roots, $\eta_{\a_i} = 1$.  The
signs can be consistently assigned to all generators since  
nonsimple roots can be written as sum of  
 two  other roots.

 The no torsion theorem  requires a $B_r$-``tail like'' structure which is
 fulfilled only by the $A_{2r}^{(2)}$ and $D_{r+1}^{(2)}$ (see appendix N of ref.
 \cite{cornwell}).  In both cases
 the automorphism is of order 2 (i.e. $ \s ^2 =1 $). 
 Let us denote by $\a$ the roots of the untwisted algebra $\lie$.  
 For the $A_{2r}^{(2)}$ case, the automorphism is defined by 
 \be
 \s ({\a_1}) = {\a_{2r}}, \;\; \s ({\a_2}) = {\a_{2r-1}}
\;\; \cdots ,\s ({\a_{r-1}}) = {\a_{r}}
\label{fol-a2n}
\ee
whilst for  the $D_{r+1}^{(2)}$, the automorphism acts only in the
``fish tail'' of the Dynkin diagram of $D_{r+1}$, i.e. 
\be 
\s (E_{\a_1}) = E_{\a_{1}}, \;\;\cdots ,\s (E_{\a_{r-1}}) = E_{\a_{r-1}},
\;\;\s (E_{\a_{r}}) = E_{\a_{r+1}}
\label{fol-dn}
\ee
The automorphism $\s$  defines a decomposition of  the algebra $\lie = \lie _{even}
\bigcup \lie _{odd}$. The twisted affine algebra is
then constructed from $\lie $ assigning an affine index $m \in Z$
 to the generators in $\lie _{even}$ while $m \in Z + {1\o 2}$ to those in 
 $\lie _{odd}$ (see appendix N of \cite{cornwell}).
 
The simple root step operators for $A_{2r}^{(2)}$     are  
\be
E_{\b_i} = E^{(0)}_{\a_i} + E^{(0)}_{\a_{2r-i+1}}, \;\; i= 1, \cdots ,r 
 \;\;
E_{\b_0}^{({1\o 2})} =  E^{({1\o 2})}_{-\a_1 - \cdots - \a_{2r}} 
\label{simplea2n}
\ee
corresponding to the 
  simple and highest roots
 \be
\b_i = {1\o 2} (\a_i + \a_{2r-i+1}) \;\; i= 1, \cdots ,r,  \;\; \;\; 
 -\a_0 =  \a_1 + \cdots + \a_{2r} = 2(\b_1 + \cdots \b_r)
 \label{simprootsa2n}
 \ee
  respectively. For $D_{r+1}^{(2)}$, the simple root step operators are
\br
E_{\b_i} = E^{(0)}_{\a_i},\; i= 1, \cdots ,r-1, \;\quad
E_{\b_r} = E^{(0)}_{\a_r} + E^{(0)}_{\a_{r+1}} ,\; \nonu\\
E_{\b_0}^{({1\o 2})} =  E^{({1\o 2})}_{-\a_1 - \cdots - \a_{r-1} - \a_{r+1}}-
E^{({1\o 2})}_{-\a_1 - \cdots - \a_{r-1} - \a_{r}}
\label{simpledn}
\er
corresponding to the 
  simple and highest roots
 \be
\b_i =  \a_i \;\; i= 1, \cdots ,r-1,  \;\; \;\;
\b_r = {1\o 2} (\a_r + \a_{r+1}), \quad
-\a_0 = \a_1 + \cdots \a_{r-1} + {1\o 2}(\a_{r} + \a_{r+1})=  
 \b_1 + \cdots \b_r
 \label{simplrootsdn}
 \ee
 where have denoted by $\b$ the roots of the twisted (folded) algebra.
 
 The corresponding  torsionless affine NA Toda models are defined by introducing the following grading operators :
 \be
 Q_{A_{2r}^{(2)}} = 2(2r-1)\hat d + \sum_{i \neq r, r+1}^{2r}{{2\lambda_i \cdot H}\o {\a_i^2}},\quad\quad\quad
 Q_{D_{r+1}^{(2)}} = (2r-2)\hat d + \sum_{i=1}^{r}{{2\lambda_i \cdot H}\o {\a_i^2}}
 \label{qdn}
 \ee
 for  $A_{2r}^{(2)}$ and $D_{r+1}^{(2)}$ respectively, where $\lambda_i $ are
 the fundamental weights of the untwisted algebra $\lie $, i.e. ${{2\lambda_i
 \cdot \a_j }\o {\a_j^2}} = \d_{ij}$.
 Both models are specified by the constant grade $\pm 1$ operators $\hat{\eps}_{\pm}$
 \be
 \hat {\eps}_{\pm} = \sum_{i=1}^{r-2} c_{\pm i}E_{\pm \b_i} + c_{\pm (r-1)}E_{\pm
 (\b_{r-1}+\b_r)} + c_{\pm r} E_{\mp \b_0}^{(\pm {{1\o 2}})}
 \label{eps}
 \ee
 where $\b_i$ are the simple roots of the twisted affine algebra specified in
 (\ref{simprootsa2n}) and  in (\ref{simplrootsdn}).
 
 The grading operators (\ref{qdn}) determine the zero
 grade subalgebra  in both cases to be $\lie_0 = 
 SL(2)\otimes U(1)^{r-1}$ generated
 by $ {E^{(0)}_{\pm \b_r}, h_1, \cdots , h_r } $. 
  Hence the zero grade subgroup is 
 parametrized as in (\ref{63}) with $\eta =0= \nu $ .  The  
  factor group $\lie /{\lie_0^0}$
  is given in
  (\ref{63a}), where
 $\lie_0^0 $ is generated by 
 $Y\cdot H = (\frac{2\mu_{r}}{\b_{r}^{2}}
-\frac{2\mu_{r-1}}{\b_{r-1}^{2}})\cdot H$ and    $\mu_i$ are the fundamental
 weights  of the twisted algebra i.e. ${{2\mu_i \cdot \b_j} \o {\b_j^2}}=
 \d_{ij}$
 . In order to decouple the $\varphi_i, \;\; i=1, \cdots , r-1$ 
 we choose an orthonormal basis for the Cartan subalgebra( i.e.
  $\Phi (H) = {\cal H}_i \varphi_i  $)
where
\be 
{\cal H}_i  = ( \a_i + \cdots \a_{2r-i+1})\cdot H, \;\;
   Y\cdot H = {\cal H}_r,  \;\; 
 Tr ({\cal H}_i{\cal H}_j) = 2\d_{ij}, \; i,j = 1, \cdots r
\label{cartana2n}
\ee
 and 
\be 
 {\cal H}_i = (\a_{r-i+1}  + \cdots + \a_{r+1})\cdot H, \;\;
  Y\cdot H = {\cal H}_r,   \;\;
 Tr ({\cal H}_i{\cal H}_j) = \d_{ij}, \; i,j = 1, \cdots r
\label{cartandn}
\ee
for   $A_{2r}^{(2)}$ and $D_{r+1}^{(2)}$ respectively.

The Lagrangeans of the corresponding (axial gauged IMs), based on the twisted affine algebras in consideration, are obtained 
from (\ref{action}) (with $\eta = 0$)  and the $\eps_{\pm}$ given by (\ref{eps}). Up to a multiplicative
factor$ -{k \o {2\pi}}$ they are given by 
\be
{\cal L}_{A_{2r}^{(2)}} = {{\pa \chi \bar \pa \psi }\o {1+ {1\o 2}\psi \chi }}
+ {1\o 2} \sum_{i=1}^{r-1}\pa \varphi_i \bar \pa  \varphi_i - V_{A_{2r}^{(2)}}
\label{acta2n}
\ee
and
\be 
{\cal L}_{D_{r+1}^{(2)}} = 2{{\pa \chi \bar \pa \psi }\o {1+ \psi \chi }}
+ {1\o 2} \sum_{i=1}^{r-1}\pa \varphi_i \bar \pa  \varphi_i - V_{D_{r+1}^{(2)}}
\label{actdn}
\ee
The potentials of these twisted affine singular NA - Toda models have the form
\be
V_{A_{2r}^{(2)}} = \sum_{i=1}^{r-2}|c_i|^2 e^{-\varphi_i + \varphi_{i+1}} +
{1\o 2}|c_r|^2 e^{2\varphi_1} + |c_{r-1}|^2e^{-\varphi_{r-1}}(1+ \psi \chi )
\label{pota2n}
\ee
and 
\be
V_{D_{r+1}^{(2)}} = \sum_{i=1}^{r-2}|c_i|^2 e^{-\varphi_i + \varphi_{i+1}} +
{1\o 2}|c_r|^2 e^{\varphi_1} + |c_{r-1}|^2e^{-\varphi_{r-1}}(1+ 2\psi \chi )
\label{potdn}
\ee
The T -selfdual models described by  (\ref{affaction}), (\ref{acta2n}) and (\ref{actdn}) turns out
to coincide with those proposed by Fateev in \cite{Fat}.

\sect{Zero Curvature}
The equations of motion for the NA Toda models are known to be of the form
\cite{leznov-saveliev}
\be 
\bar \pa (B^{-1} \pa B) + [\hat {\eps_-}, B^{-1} \hat {\eps_+} B] =0, 
\quad \pa (\bar \pa B B^{-1} ) - [\hat {\eps_+}, B\hat {\eps_-} B^{-1}] =0
\label{eqmotion}
\ee
 The subsidiary constraint $J_{Y \cdot H} = Tr(B^{-1} \pa B Y\cdot H)=
  \bar J_{Y \cdot H} = Tr(\bar \pa B B^{-1}Y \cdot H )=        0$ can be
 consistenly imposed  since $[Y\cdot H, \hat {\eps_{\pm}}]=0$ (as can be 
 seen from
 (\ref{eqmotion}) by taking the trace with $Y.H$). We next consider the 
  axial models only. As we have mentioned in Sect.3 the vector models
  Lagrangeans (and equations of motion) can be obtained from the axial ones by
  the nonlocal change of the fields (see for example eqs.(\ref{3.9})
  and (\ref{3.10}) for the $A_r^{(1)}$ and $B_r^{(1)}$ models). 
 Solving those equations 
 for the nonlocal
 field $R$ yields 
 \be
 \pa R = ({{Y\cdot \a_r}\o {Y^2}}) {{\psi \pa \chi }\o \Delta }e^{\Phi(\a_r)},
 \;\; \;\;\; 
\bar \pa R = ({{Y\cdot \a_r}\o {Y^2}}) {{\chi \bar \pa \psi }\o \Delta }e^{\Phi(\a_r)}
\label{bh}
\ee
The equations of motion for the fields $\psi, \chi $ and $\varphi_i, i=1,
\cdots , r-1$  (obtained from (\ref{eqmotion})  by imposing the constraints
(\ref{bh})) coincide precisely with the Euler-Lagrange equations derived from
(\ref{acta2n}) and (\ref{actdn}).  Alternatively, (\ref{eqmotion}) admits a
zero curvature representation \\ 
$\pa \bar A - \bar \pa A + [A, \bar A] =0$, where
\be
A= \hat {\eps_-} + B^{-1} \pa B,\quad  
\bar A= -B^{-1}\hat {\eps_+}  B
\label{zcc}
\ee
Whenever the constraints (\ref{bh}) are incorporated into $A$ and $\bar A$ in
(\ref{zcc}), equations (\ref{eqmotion}) yields the zero curvature
representation of the affine {\it singular}  NA -Toda models.  Such argument is valid for all conformal, affine and conformal affine
NA - Toda models, in particular for the torsionless class of models discussed in
the previous section.

Using the explicit parametrization of $B$ given in (\ref{632}),(with $\eta = 0$)  the
corresponding $\hat {\eps_{\pm}}$ specified in (\ref{eps}), 
 (\ref{simprootsa2n}) and (\ref{simplrootsdn}) together with  
(\ref{bh})(with $Y$ given by eqs.
 (\ref{cartana2n}) and (\ref{cartandn})), we obtain  the flat connections $A$ and $\bar A$ in the following form :

(a) the $ A_{2r}^{(2)}$ affine NA - Toda model 
\br
A_{A_{2r}^{(2)}} &=& \sum_{i=1}^{r-2} c_i (E^{(0)}_{-\a_i} +
E^{(0)}_{-\a_{2r-i+1}}) +  
 c_{r-1} (E^{(0)}_{-\a_r - \a_{r-1}} + E^{(0)}_{-\a_{r+1}-\a_{r+2}})\nonu \\ 
  &+&
 c_r  E^{(-{1\o 2})}_{ \a_1 + \cdots + \a_{2r}}
 + 
 \pa \psi e^{-{1\o 2}R}(E^{(0)}_{\a_r} + E^{(0)}_{\a_{r+1}}) 
+\sum_{i=1}^{r-1} \pa \varphi_i {\cal H}_i \nonu \\
&+& 
{{\pa \chi }\o \Delta }e^{{1\o 2}R}(E^{(0)}_{-\a_r} + E^{(0)}_{-\a_{r+1}})
\label{zcca2n}
\er
and 
\br
-\bar A_{A_{2r}^{(2)}} &=& \sum_{i=1}^{r-2} c_i e^{-\varphi_i + \varphi_{i+1}}
(E^{(0)}_{\a_i} +
E^{(0)}_{\a_{2r-i+1}}) + c_r e^{2\varphi_1}
 E^{({1\o 2})}_{ -\a_1 - \cdots - \a_{2r}}\nonu \\ 
&+& c_{r-1}e^{-\varphi_{r-1}}(E^{(0)}_{\a_r + \a_{r-1}} +
 E^{(0)}_{\a_{r+1}+\a_{r+2}})\nonu \\
&+&c_{r-1}\psi e^{-{1\o 2}R -\varphi_{r-1}}(E^{(0)}_{\a_{r+1}+\a_r + \a_{r-1}} -
 E^{(0)}_{\a_r+ \a_{r+1}+\a_{r+2}})\nonu \\ 
&+& c_{r-1}\chi e^{{1\o 2}R -\varphi_{r-1}}(E^{(0)}_{ \a_{r-1}} -
 E^{(0)}_{\a_{r+2}}) 
+ c_{r-1}\psi \chi e^{-\varphi_{r-1}}(E^{(0)}_{\a_r + \a_{r-1}} -
 E^{(0)}_{ \a_{r+1}+\a_{r+2}}) \nonu \\ 
&+& {1\o 2}c_{r-1}\psi ^2 \chi e^{-\varphi_{r-1}-{1\o 2}R}
(E^{(0)}_{\a_{r+1}+\a_r + \a_{r-1}} -
 E^{(0)}_{\a_r+ \a_{r+1}+\a_{r+2}}) 
\label{zcca2nbar}
\er

(b) the $D_{r+1}^{(2)}$ affine NA - Toda model

\br
A_{D_{r+1}^{(2)}} &= & \sum_{i=1}^{r-2} c_i E^{(0)}_{-\a_i} 
 +  
 c_{r-1} (E^{(0)}_{-\a_r - \a_{r-1}} + E^{(0)}_{-\a_{r-1}-\a_{r+1}})\nonu \\ 
  &+&
 c_r  (E^{(-{1\o 2})}_{ (\a_1 + \cdots + \a_{r-1}+ \a_{r+1})} -
 E^{(-{1\o 2})}_{ (\a_1 + \cdots + \a_{r}+ \a_{r+1})})
 + 
 \pa \psi e^{-{1\o 2}R}(E^{(0)}_{\a_r} + E^{(0)}_{\a_{r+1}}) \nonu \\
&+&\sum_{i=1}^{r-1} \pa \varphi_i {\cal H}_i 
+ 
{{\pa \chi }\o \Delta }e^{{1\o 2}R}(E^{(0)}_{-\a_r} + 
E^{(0)}_{-\a_{r+1}})
\label{zccdn}
\er
and 
\br
-\bar A_{D_{r+1}^{(2)}} &=& \sum_{i=1}^{r-2} c_i e^{-\varphi_i + \varphi_{i+1}}
E^{(0)}_{\a_i}   
+ c_{r-1}e^{-\varphi_{r-1}}(E^{(0)}_{\a_r + \a_{r-1}} +
 E^{(0)}_{\a_{r+1}+\a_{r-1}})\nonu \\
&+& 2c_{r-1}\psi e^{-{1\o 2}R -\varphi_{r-1}}E^{(0)}_{\a_{r+1}+\a_r
+\a_{r-1} }
+ 2c_{r-1}\chi e^{{1\o 2}R -\varphi_{r-1}}E^{(0)}_{ \a_{r-1}}\nonu \\  
&+& 2c_{r-1}\psi \chi e^{-\varphi_{r-1}}(E^{(0)}_{\a_{r+1} + \a_{r-1}} 
 + E^{(0)}_{\a_{r-1} + \a_{r}}) 
+ c_{r-1}\psi ^2 \chi e^{-{1\o 2}R -\varphi_{r-1}}
E^{(0)}_{\a_{r+1}+\a_r + \a_{r-1}} \nonu \\
 &+& c_{r+1}e^{\varphi_1}
(E^{({1\o 2})}_{ -(\a_1 + \cdots + \a_{r-1}+ \a_{r+1})} -
 E^{({1\o 2})}_{ -(\a_1 + \cdots + \a_{r}+ \a_{r+1})}) 
\label{zccdnbar}
\er
For the untwisted affine $B^{(1)}_r$ model of the previous section the zero
curvature representation is obtained from the following flat connections :
\br
A_{B_{r}^{(1)}} &= & \sum_{i=1}^{r-2} c_i E^{(0)}_{-\a_i}
+  
 c_{r-1} E^{(0)}_{-\a_r - \a_{r-1}} 
+c_r  E^{(-{1})}_{ \a_1 +2(\a_2 + \cdots +  \a_{r})}\nonu \\  
&+& \pa \psi e^{-{1\o 2}R}E_{\a_r}^{(0)} 
+\sum_{i=1}^{r-1} \pa \varphi_i {\cal H}_i 
+{{\pa \chi }\o \Delta }e^{{1\o 2}R } E_{-\a_r}^{(0)}
\label{zccbn}
\er
\br
-\bar A_{B_{r}^{(1)}} &=& \sum_{i=1}^{r-2} c_i e^{-\varphi_i + \varphi_{i+1}}
E^{(0)}_{\a_i}    + c_r e^{\varphi_1 + \varphi_2 }
E^{(1)}_{ -(\a_1 +2(\a_2 + \cdots +  \a_{r}))} 
+ 2 \chi e^{\varphi_{r-1}+ {1\o 2}R}E_{\a_{r-1}}^{(0)} \nonu \\
&+& c_{r-1}(1+2 \psi \chi )
e^{-\varphi_{r-1}}E^{(0)}_{\a_{r-1} + \a_r} 
-2c_{r-1}e^{-\varphi_{r-1} -
{1\o 2}R} \psi (1+ \psi \chi )E^{(0)}_{\a_{r-1} 
+ 2\a_r}
\label{zccbnbar}
\er
The zero curvature representation of the  subclass of torsionless singular affine NA - Toda
models shows that they are in fact classically integrable field theories. 
The  construction of the previous sections provides a  systematic affine Lie
algebraic structure underlying those models, which is known to play the crucial role in the
construction of their finite energy soliton solutions.

\sect{Conclusions}

We have constructed a class of affine (and conformal affine) NA-Toda models from the gauged two-loop 
WZNW models, in which left and right symmetries are incorporated by a suitable 
choice of grading operator $Q$ and of grade $\pm 1$ 
constant generators $\eps_{\pm}$. We have shown that for non abelian zero grade subalgebra $\lie_0$,
 it is possible to reduce even further the phase space by constraining to
zero the currents associated to generators $Y\cdot H$, 
commuting with $\eps_{\pm}$  ($ Y\cdot H \in \lie_0^0)$). 
 There exists two  inequivalent manners to gauge fix
$\lie_0^0 = U(1)$-the { axial and the vector} gaugings. Similarly to the T-duality
transformations between the axial and vector gauged G/H - WZNW models, one can find the
{\it off-critical} counterpart of the {\it conformal } T-duality, relating now
the axial and vector families of IM's constructed in Sect. 2 . We further
analize the problem of deriving  the Lie algebraic condition 
 which defines a class of {\it T-selfdual
torsionless models}, for the case $\lie_0^0 = U(1)$.  The action  
for those models were systematicaly 
constructed and shown to coincide with those 
proposed by Fateev \cite{Fat}, describing the {\it strong coupling limit} of specific 2-d models 
representing the complex 
sine-Gordon (i.e. Lund-Regge \cite{lund}) interacting with Toda-like models. Their {\it weak coupling limit}
appears to be  the Thirring model coupled to certain affine Toda theories 
\cite{Fat}.

As we have mentioned in sects. 1 and 3 the conformal $\sigma$-model limits of the axial and vector
dyonic IM's describe critical  $D=r+1$ strings on black hole backgrounds \cite{ginsparg}. One may
wonder whether the {\it nonconformal} (i.e. off-critical) dyonic IM's
 (representing {\it integrable
perturbations} of these string models ) have some string field theory
 applications. As is well known
\cite{zamol} the relation between 2-d conformal model and its 
integrable perturbations allows to
describe the off-critical behaviour of the original conformal model 
as well as the RG flow from
ultraviolet to infrared (in the case of unitary CFT's ) see \cite{haa}, \cite{ts} and references therein. 
 Hence the properties of the admissible
integrable perturbations of the conformal $\sigma $-models  are 
an important ingredient in the description
of the space of string backgrounds (satisfying certain low energy physical requirements). As it advocated by B. Zwiebach 
\cite{zwie} the {\it nonconformal} 
versions of the conformal string
backgrounds appears to be the main tool for the construction of the {\it off-shell string field theory
(SFT) }. Among all the possible
nonconformal backgrounds the {\it integrable }ones 
(admiting "worldsheet" soliton solutions) have the
advantage to offer powerfull methods for studying the string S-duality. 

The dyonic integrable models studied in the present paper represent the simplest family
 of IM's with one
U(1) global symmetry. Translated to the string language this means that the target-space metrics 
 ($E_{MN}(X), \phi( X), T(X)$) are independent of one of 
 the coordinates \\
 $ X_M $ ($\theta={1\o 2} ln {\chi
 \o \psi} $ in our case) and that the relevant
 operator does not break this isometry. Models with more isometries (generic abelian T-duality) or
 those admiting isotropies ( nonabelian T-duality) appears to be useful for the construction of
 physically intersting string models (\cite {giveon}). 
 The problem of construction of
 their integrable perturbations requires to consider more general
  affine $\hat\lie_r$-NA Toda models
 defined by grading operators as for example $Q_{a,b,\cdots }
= h_{a,b, \cdots } \hat {d} + 
\sum_{i\neq a,b \cdots }^{n}\frac{2\lambda_{i}\cdot H}{\alpha_{i}^{2}}$ and  appropriately choosen
$\epsilon_{\pm}$. They should allow larger(than U(1)) invariant subgroup $\lie_0^0 $
 (non-abelian in general). An important characteristic of such IM's is that their physical fields belongs to , say
$ \lie_0 /{\lie_0^0} = 
{{SL(2)\otimes U(1)^{r-1}}\o {U(1)^{s}}}, \\
\quad s=2, \cdots r-1$ (i.e  with s isometries), or  
$ \lie_0 /{\lie_0^0} = 
{{SL(2)\otimes SL(2)\otimes U(1)^{r-2}} \o {U(1)^{s}}}$, or  
$ \lie_0 /{\lie_0^0} = 
{{SL(3)\otimes U(1)^{r-2}}\o {U(2)}}$ (the corresponding IM's  represent string backgrounds with one
isometry and SL(2) as isotropy group ) ,etc. 
The methods needed in the construction of such IM's and for the investigation of their T-duality
properties appears to be straightforward generalization of the methods we have developed for the
simplest case of $\lie_0^0 = U(1)$ . An intersting {\it open problem }is the classification of the affine
NA-Toda IM's according to the number of the physical fields ( i.e. the dimension of the string target
space ) and their symmetry groups $\lie_0^0 $ ( i.e. the symmetries of the string backgrounds). We
hope that the simplest dyonic IM's studied in the present paper and the methods we have developed might
contribute to the complete description of the space of the integrable (nonconformal) string
backgrounds.

{\bf Acknowledgments}  We are grateful to CNPq, FAPESP and UNESP for 
financial support.

\end{document}